# Mahaux-Weidenmüller approach to cavity quantum electrodynamics and complete resonant down-conversion of the single photon frequency

M. Sumetsky

Aston Institute of Photonic Technologies, Aston University, Birmingham B4 7ET, United Kingdom

*m.sumetsky@aston.ac.uk*

It is shown that a broad class of cavity quantum electrodynamics (QED) problems – which consider the resonant propagation of a single photon interacting with quantum emitters (QEs), such as atoms, quantum dots, or vacancy centers – can be solved directly without application of the second quantization formalism. In the developed approach, the Hamiltonian is expressed through the ket-bra products of collective (photon + cavities + QEs) states. Consequently, the S-matrix of input-output problems is determined exactly by the Mahaux-Weidenmüller formula, which dramatically simplifies the analysis of complex cavity QED systems. First, this approach is illustrated for the problem of propagation of a photon resonantly interacting with $N$ two-level QEs arbitrary distributed inside the optical cavity. Solution of this problem manifests the effect of cumulative action of QEs previously known for special cases. Can a similar cumulative action of QEs enhance the inelastic resonant transmission of a single photon? We solve this problem for the case of an optical cavity having two modes resonantly coupled to electronic transitions of $N$ three-level QEs. It is shown that the described structure is the simplest realistic structure which enables the down-conversion of the single photon frequency with the amplitude approaching unity in the absence of the external driving field and sufficiently small cavity losses and QE dissipation. Overall, the simplicity and generality of the developed approach suggest a practical way to identify and describe new phenomena in cavity QED.

## I.    INTRODUCTION

The solution of the majority of equations describing complex quantum systems cannot be found without simplifications based on additional physical assumptions. For variety of systems, the simplification consists in selecting a relatively small number of states, which the system can populate during the process under interest, while ignoring all others. For the input-output problems commonly considered in quantum electrodynamics (QED) [1-3], this may happen when the energy of the incident particle is close to the eigenenergies of one or more localized states of the system. For example, Fig. 1(a) illustrates a system of two semi-infinite waveguides weakly coupled to a single state $|\Psi_1\rangle$ of the cavity. If the eigenenergy $\hbar\Omega_1^{(0)}$ of state $|\Psi_1\rangle$ is close to the input particle energy $\hbar\Omega$ and separated from other eigenenergies, then the expression for the resonant transmission amplitude from waveguide 1 to waveguide 2 is reduced to the Breit-Wigner formula [4, 5]:

$$S_{12} = \frac{2\pi i W_{11} W_{12}}{\Omega - \Omega_1^{(0)} + i(\pi W_{11}^2 + \pi W_{12}^2)} \; .$$   (1)

Here $W_{1p}$, $p = 1, 2$, are coupling parameters between localized state $|\Psi_1\rangle$ and waveguide states $|\Lambda_{p,\Omega}\rangle$ and the imaginary part of the eigenenergy, $\mathrm{Im}\,\Omega_1^{(0)}$, takes into account losses caused by the interaction with the environment not included in the model of Fig. 1(a). Equation (1) was used to describe a wide range of phenomena in nuclear, atomic, and molecular scattering [4-7], propagation of electrons in superlattices and microstructures [8-13], and propagation of photons in cavity QED [1, 3, 14-16].

Remarkably, Eq. (1) includes just a few parameters describing a quantum system which may be quite complex. It is also notable that this equation can describe the elastic as well as inelastic processes. While for the elastic transmission, the energy $\hbar\Omega$ of the propagating quantum particle is conserved, it can change in the process of inelastic transmission. In order to apply Eq. (1) to the case of inelastic transmission, we assume that states $|\Psi_1\rangle$ and $|\Lambda_{p,\Omega}\rangle$ are the *collective* states of the ingoing particle and other particles interacting with this particle and forming a closed system all together. Then the total energy $\hbar\Omega$ of all interacting particles is conserved again, while these particles can exchange energy in the process of propagation.

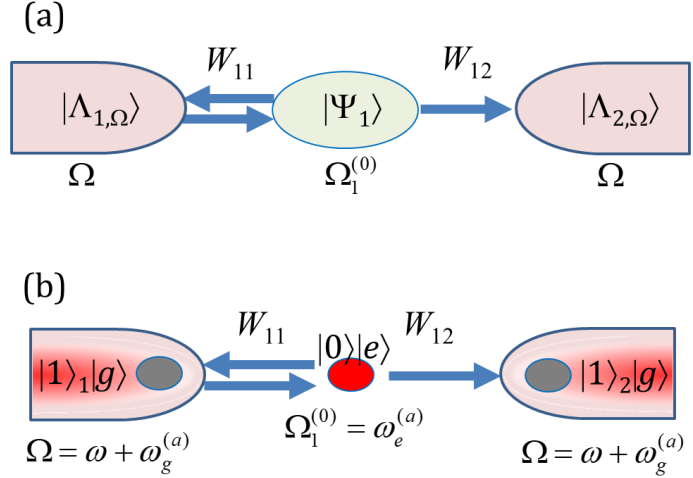

(a)

(b)

**Fig. 1.** (a) Two waveguide states $\left|\Lambda_{1,\Omega}\right\rangle$ and $\left|\Lambda_{2,\Omega}\right\rangle$ coupled to cavity state $\left|\Psi_1\right\rangle$. (b) Formulation of the input-output problem in terms of collective states of a photon and a QE. Here the waveguide states $\left|\Lambda_{1,\Omega}\right\rangle = \left|1\right\rangle_1 \left|g\right\rangle$ and $\left|\Lambda_{2,\Omega}\right\rangle = \left|1\right\rangle_2 \left|g\right\rangle$ are the collective states of an input and output photon and a QE in the ground state. The localized state $\left|\Psi_1\right\rangle = \left|0\right\rangle\left|e\right\rangle$ is the collective state of a QE in the excited state and no photon.

Consider, for example, a single photon resonantly propagating through a two-level quantum emitter (QE) positioned in between two waveguides [Fig. 1(b)]. Here and below a QE is defined as an elementary quantum system, such as an individual atom, quantum dot, or a vacancy center, which is capable of optical transitions with emission or absorption of a single photon. Assume that, originally, the QE is in the ground state $\left|g\right\rangle$ with energy $\hbar\omega_g^{(a)}$ and the energy of the input photon $\hbar\omega$ is close to the difference $\hbar\omega_e^{(a)} - \hbar\omega_g^{(a)}$ between the energies of the excited and ground states of QE, $\left|e\right\rangle$ and $\left|g\right\rangle$. We define the input and output collective states of the photon and QE as $\left|\Lambda_{p,\Omega}\right\rangle = \left|1\right\rangle_p \left|g\right\rangle$ where $\left|1\right\rangle_p$ is the state of the photon with energy $\hbar\omega$ in waveguide $p$. The total energy of these states is $\hbar\Omega = \hbar\omega + \hbar\omega_g^{(a)}$. We define the collective localized state (excited QE and no photon) as $\left|\Psi_1\right\rangle = \left|0\right\rangle\left|e\right\rangle$. The eigenenergy of this state is $\hbar\Omega_1^{(0)} = \hbar\omega_e^{(a)}$. The coupling parameters $W_{1p}$ in this equation are determined by the amplitudes of transition $\left|g\right\rangle \leftrightarrow \left|e\right\rangle$ induced by the evanescent fields of waveguides [17], while $\mathrm{Im}\,\Omega_1^{(0)}$ is the dissipation rate of the excited state of QE. For relatively small dissipation, $\mathrm{Im}\,\Omega_1^{(0)} << W_{1p}^2$, and equal coupling, $W_{11} = W_{12}$, the transmission amplitude $S_{12} = 1$. This means that a single QE can support the full transmission of a single photon between weakly coupled waveguides with the probability equal to unity. This result is similar to that obtained previously [14, 15].

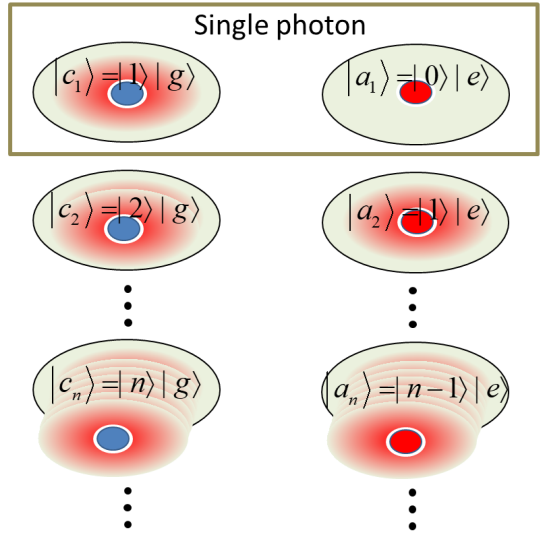

**Fig. 2.** Illustration of eigenfunctions of the Hamiltonian $H$ of identical photons and a two-level QE in an optical cavity defined by Eq. (2). Each of these eigenfunctions is the collective state of several identical photons localized in the cavity and a QE in its ground or excited state. The rectangle outlines two states, which should only be taken into account for the resonant transmission of a single photon.

The above arguments, which validate the application of the Breit-Wigner formula, Eq. (1), to the problem of resonant propagation of a single photon, are not based on the second quantization formalism. These arguments suggest significant simplification of equations commonly used to describe the resonant propagation of a single photon and, presumably, a few photons. Instead of separating the Hamiltonian of the system into the part describing the photons, which is expressed through the products of creation and annihilation operators, we can present it in the standard form as $H = \sum_n \hbar\Omega_n |\Psi_n\rangle\langle\Psi_n|$ where $|\Psi_n\rangle$ and $\hbar\Omega_n$ are the eigenfunctions and eigenenergies of $H$. Using this presentation, we can simplify the Hamiltonian by excluding all eigenfunctions, whose eigenenergies are not in resonance with the energy $\hbar\Omega$ of the input state. To illustrate this approach, consider the Hamiltonian of identical photons with energies $\hbar\omega^{(c)}$ and a two-level QE in an optical cavity:

$$H = \hbar\omega^{(c)}b^{\dagger}b + \hbar\omega_g^{(a)}|g\rangle\langle g| + \hbar\omega_e^{(a)}|e\rangle\langle e|$$
$$= \sum_n \left( \hbar\Omega_n^{(c)}|c_n\rangle\langle c_n| + \hbar\Omega_n^{(a)}|a_n\rangle\langle a_n| \right) \qquad (2)$$

The first line in this equation presents $H$ in terms of photon creation and annihilation operators, $b^{\dagger}$ and $b$. The second line is the representation of $H$ through the ket-bra products of its eigenfunctions $|c_n\rangle$ and $|a_n\rangle$ illustrated in Fig. 2. The eigenfunction $|c_n\rangle = |n\rangle|g\rangle$ is the collective state of $n$ photons in the cavity and a QE in its ground state $|g\rangle$. The eigenenergy of this state is $\Omega_n^{(c)} = \hbar\omega^{(c)}n + \hbar\omega_g^{(a)}$. Similarly, $|a_n\rangle = |n-1\rangle|e\rangle$ is the collective state of $n-1$ photons in the cavity and QE in its excited state $|e\rangle$. The eigenenergy of this state is $\Omega_n^{(a)} = \hbar\omega^{(c)}(n-1) + \hbar\omega_e^{(a)}$. If only a single photon participates in the problem of our interest, all the terms in the second line of Eq. (2) can be neglected except for two terms with $n = 1$ outlined by the rectangle in Fig. 2. Then the Hamiltonian in Eq. (2) is reduced to

$$H_1 = \hbar\Omega_1^{(c)}|c_1\rangle\langle c_1| + \hbar\Omega_1^{(a)}|a_1\rangle\langle a_1|$$
$$= \hbar(\omega^{(c)} + \omega_g^{(a)})|1\rangle|g\rangle\langle g|\langle 1| + \hbar\omega_e^{(a)}|0\rangle|e\rangle\langle e|\langle 0|. \qquad (3)$$

Here we note that the approach based on the reduction of the Hamiltonian expressed through the creation and annihilation operators [first line in Eq. (2)] to the Hamiltonian expressed only through its resonant eigenfunctions (Eq. (3)) is quite general. In this paper, we develop this approach in application to the problem of resonant propagation of a single photon through complex structures of optical cavities and QEs. This problem is reduced to the input-output problem considering several localized states coupled to several waveguides illustrated in Fig. 3, which, as was shown by Mahaux and Weidenmüller in 1969, can be solved exactly [18-20]. The celebrated Mahaux-Weidenmüller formula introduced in Sec. II is the generalization of the Breit-Wigner formula, Eq. (1), to the case of multiple resonant input-output and localized states. It has been applied previously to a range of problems in nuclear scattering [18-21], conductance of microelectronic and nanoelectronic devices [22-24], and transmission of photonic microstructures [25-28]. However, the applications of this formula to the cavity QED problems – leading to their analytical solutions – have not been thoroughly explored. The reason is presumably in the fact that, conventionally, the QED Hamiltonians are presented through the creation-annihilation operators [Eq. (2), line one] rather than their eigenfunctions [Eq. (2), line two]. The latter presentation, however, was the major assumption of the Mahaux-Weidenmüller theory.

This article is organized as follows. After introducing the Mahaux-Weidenmüller formalism in Sec. II, we proceed with its applications. To this end, we reformulate the problems of resonant propagation of a single photon in terms of the eigenfunctions of corresponding Hamiltonians. Similar to the example considered above, these eigenfunctions represent the collective states of a photon and QEs. In Sec. III, we consider the resonant propagation of a single photon interacting with two-level QEs in an optical cavity [14, 15, 29-36]. This classical problem is reduced to the consideration of $N+1$ localized collective states and two semi-infinite waveguide collective states. We show that, for this case, the Mahaux-Weidenmüller formula yields the expressions for the S-matrix generalizing those obtained previously using special forms of QED Hamiltonians. In particular, we show that the resonant transmission amplitude of a single photon can approach unity due to the cumulative action of QEs. This result points us to the question of whether a similar cumulative action of QEs can enhance the *inelastic* transmission of a single photon and, in particular, ensure its complete frequency conversion. In Sec. IV, we address this question by considering the problem of inelastic resonant transmission of a single photon through an optical microcavity with two eigenfrequencies resonantly coupled to three levels of $N$ QEs. We determine the conditions of complete frequency conversion of a single photon and analyze the effect of cumulative action of QEs in this process. In Sec. V, we discuss and summarize the results obtained.

## II.    MAHAUX-WEIDENMÜLLER FORMALISM

The Mahaux-Weidenmüller theory [18-20] considers a general multiparticle system which includes resonantly coupled $N$ localized collective states $\left| \Psi_n \right\rangle$ and $P$ waveguide collective states $\left| \Lambda_{p,\Omega} \right\rangle$, illustrated in Fig. 3. The localized states $\left| \Psi_n \right\rangle$, $n=1,2,...,N$, if assumed uncoupled, have complex eigenfrequencies $\Omega_n^{(0)}$. The waveguide states $\left| \Lambda_{p,\Omega} \right\rangle$ with numbers $p=1,2,...,P$ are confined along all directions in the multiparticle configuration space except for the those which correspond to the propagation of particles along the waveguide $p$. The Hamiltonian of this system is [18-20]

$$
\begin{aligned}
H = \hbar \Bigg[ &\sum_n \Omega_n^{(0)} \left| \Psi_n \right\rangle \left\langle \Psi_n \right| + \sum_p \int d\Omega \, \Omega \left| \Lambda_{p,\Omega} \right\rangle \left\langle \Lambda_{p,\Omega} \right| \\
&+ \sum_{m,n} V_{mn} \left| \Psi_m \right\rangle \left\langle \Psi_n \right| + \sum_{n,p} \int d\Omega \left( W_{np} \left| \Psi_n \right\rangle \left\langle \Lambda_{p,\Omega} \right| + H.c. \right) \Bigg],
\end{aligned}
$$

(4)

where the couplings between states and waveguides are assumed to be real [19] and direct coupling between the waveguide states is neglected. The $S$ matrix $\mathbf{S} = \left\{ S_{pq} \right\}$ of transmission amplitudes from waveguide $p$ to waveguide $q$ is expressed through the matrix of eigenfrequencies $\mathbf{\Omega}^{(0)} = \left\{ \Omega_n^{(0)} \delta_{mn} \right\}$ of uncoupled localized states $\left| \Psi_n \right\rangle$ and matrices of coupling parameters between localized states, $\mathbf{V} = \left\{ V_{nm} \right\}$, and between localized states and waveguides, $\mathbf{W} = \left\{ W_{mp} \right\}$, by the Mahaux-Weidenmüller formula [18-20]:

$$\mathbf{S} = \mathbf{I} - 2i\pi \mathbf{W}^\dagger (\Omega \mathbf{I} - \mathbf{\Omega}^{(0)} - \mathbf{V} + i\pi \mathbf{W}\mathbf{W}^\dagger)^{-1} \mathbf{W}, \tag{5}$$

where $\mathbf{I} = \left\{ \delta_{mn} \right\}$ is the unity matrix. Generally, Hamiltonian $H$ in Eq. (4) and the $S$ matrix determined by Eq. (5) describe the scattering process of several particles. In the resonance approximation, it is assumed that the separation between eigenvalues of localized states of $H$ is much smaller than the separation of these eigenvalues from the eigenvalues of localized states, which are not included in Eq. (4). For $N = 1$, the matrix element $S_{12}$ determined from this equation coincides with the Breit-Wigner formula, Eq. (1).

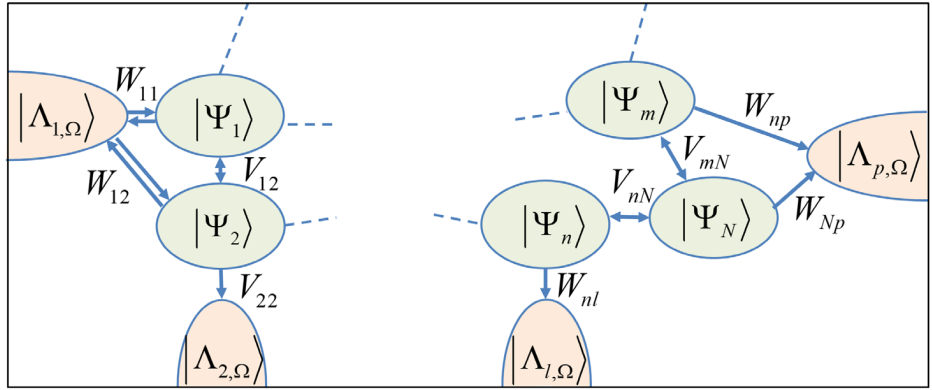

**Fig. 3.** A system of localized collective states $\left| \Psi_n \right\rangle$ and waveguide collective states $\left| \Lambda_{p,\Omega} \right\rangle$. Here $V_{mn}$ are coupling parameters between localized states and $W_{np}$ are coupling parameters between localized and waveguide states. Generally, the states are multiparticle and include collective excitations. In this paper, we consider localized and waveguide states composed of a single photon and QEs.

### III.    SINGLE EIGENFREQUENCY OPTICAL CAVITY COUPLED TO $N$ TWO-LEVEL QUANTUM EMITTERS

It is instructive to demonstrate the application of the Mahaux-Weidenmüller formalism to the classical problem of the resonant propagation of a single photon through a single eigenstate of an optical microcavity coupled to $N$ two-level QEs illustrated in Fig. 4(a) [14, 15, 29-36]. Similar to the noninteracting photon and single two-level QE discussed in the Introduction, this system can be modeled by $N+1$ localized collective eigenstates, $|c_1\rangle$ and $|a_n\rangle$, $n=1,2,...,N$, and two waveguide states, $|\Lambda_{1,\Omega}\rangle$ and $|\Lambda_{2,\Omega}\rangle$ shown in Fig. 4(b). The collective localized state $|c_1\rangle = |1\rangle|g_1\rangle|g_2\rangle...|g_N\rangle$ is composed of a single photon state $|1\rangle$ localized in the cavity and ground states $|g_n\rangle$ of $N$ QEs. The total eigenenergy of this state, if assumed uncoupled to QEs and waveguides, is $\hbar\Omega_1^{(c)} = \hbar\omega_1^{(c)} + \sum_{n=1}^{N}\hbar\omega_{g_n}^{(a)} - \frac{i}{2}\hbar\Gamma_1^{(c)}$, where $\omega_1^{(c)}$ and $\Gamma_1^{(c)}$ is the eigenfrequency and resonance width of uncoupled optical cavity and $\hbar\omega_{g_n}^{(a)}$ is the energy of the ground state of QE $n$. Localized states $|a_n\rangle = |0\rangle|g_1\rangle...|e_n\rangle...|g_N\rangle$ are the collective states of QE $n$ in its excited state $|e_n\rangle$ with complex eigenenergy $\hbar(\omega_{e_n}^{(a)} - \frac{i}{2}\Gamma_{e_n}^{(a)})$ (here $\Gamma_e^{(a)}$ is the dissipation rate of QE), other QEs remaining in their ground states, and no photon. In the resonance approximation, the direct coupling of states $|a_n\rangle$ to the waveguides is ignored. Couplings between states $|c_1\rangle$ and $|a_n\rangle$ are denoted by $V_{1n}$. For uncoupled states ($V_{1n} = 0$), the energy of $|a_n\rangle$ is $\hbar\Omega_n^{(a)} = \hbar\omega_{e_n}^{(a)} - \hbar\omega_{g_n}^{(a)} + \sum_{m=1}^{N}\hbar\omega_{g_m}^{(a)} - \frac{i}{2}\hbar\Gamma_{e_n}^{(a)}$. The collective input and output states $|\Lambda_{p,\Omega}\rangle = |1\rangle_p|g_1\rangle|g_2\rangle...|g_N\rangle$ are composed of the waveguide state of a photon with energy $\hbar\omega$ and localized ground states $|g_n\rangle$ of QEs with eigenenergies $\hbar\omega_{g_n}^{(a)}$, $n=1,2,...,N$, which are distributed inside the cavity. The total energy of these states is $\hbar\Omega = \hbar(\omega + \sum_{m=1}^{N}\hbar\omega_{g_m}^{(a)})$.

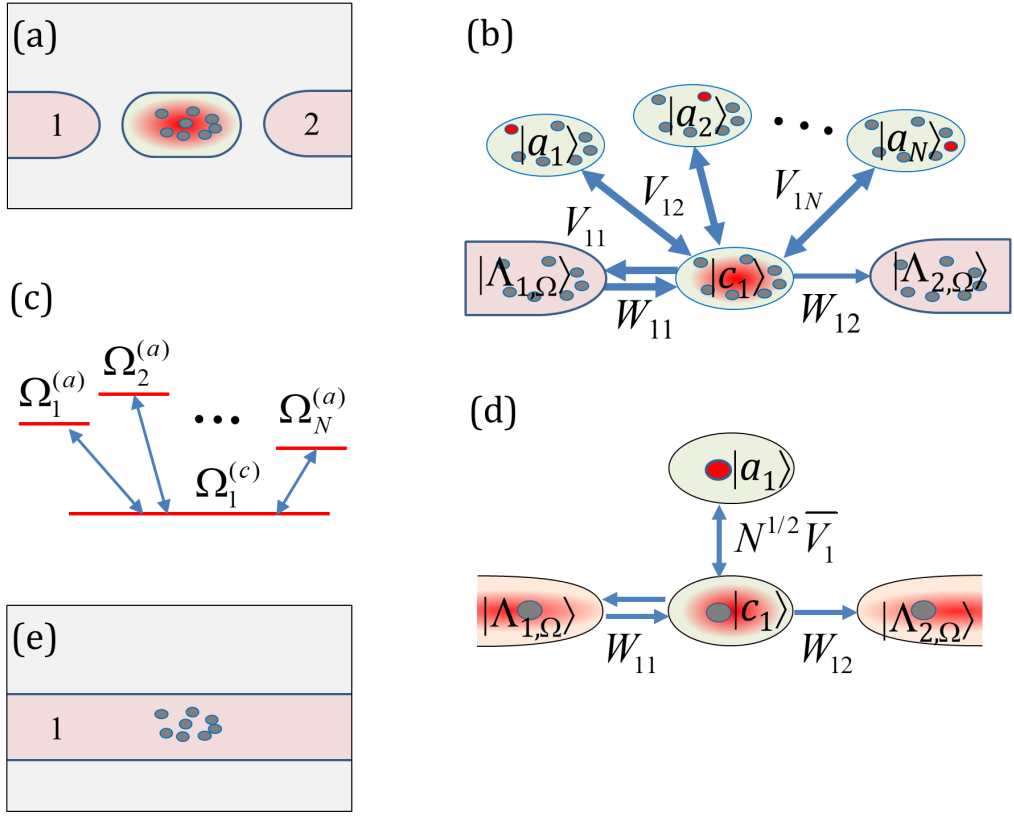

**Fig. 4.** (a) Illustration of an optical cavity with QEs coupled to the input and output waveguides. (b) Collective localized states and semi-infinite collective waveguide states, which describe the propagation of a single photon resonantly coupled to a single state of an optical cavity and $N$ two-level QEs positioned inside the cavity. (c) Schematics of transition between resonant frequencies of collective states considered, $\Omega_1^{(c)} \leftrightarrow \Omega_n^{(a)} \leftrightarrow \Omega_1^{(c)}$. (d) For identical QEs, the structure of collective states shown in (b) is reduced to the structure shown in this figure.

The $S$-matrix of the considered structure can be directly found from Eq. (5) as (see Appendix A):

$$S_{11} = 1 - 2i\pi \frac{W_{11}^2}{\Xi}, \qquad S_{12} = 2i\pi \frac{W_{11}W_{12}}{\Xi},$$

$$\Xi = \Omega - \Omega_1^{(c)} - i\pi(W_{11}^2 + W_{12}^2) - \sum_{n=1}^{N} \frac{V_{1n}^2}{\Omega - \Omega_n^{(a)}}.$$

(6)

For identical QEs arbitrary distributed in the cavity, we have $\Omega_n^{(a)} = \Omega^{(a)}$, $\omega_{g_n}^{(a)} = \omega_g^{(a)}$, and $\omega_{e_n}^{(a)} = \omega_e^{(a)}$ and Eq. (6) can be written as:

$$S_{11} = 1 - 2i\pi \frac{W_{11}^2}{\Xi}, \qquad S_{12} = 2i\pi \frac{W_{11}W_{12}}{\Xi},$$

$$\Xi = \omega - \omega_1^{(c)} + i\left(\tfrac{1}{2}\Gamma_1^{(c)} + \pi W_{11}^2 + \pi W_{12}^2\right) - \frac{N\overline{V}_1^2}{\omega + \omega_g^{(a)} - \omega_e^{(a)} + \tfrac{i}{2}\Gamma_e^{(a)}}, \qquad (7)$$

$$\overline{V}_1 = \left(\frac{1}{N}\sum_{n=1}^N \left(V_{1n}\right)^2\right)^{1/2},$$

where $\overline{V}_1$ is the root-mean-square of cavity-QE couplings $V_{1n}$. For the case of a single QE, $N=1$, and symmetric waveguide-cavity coupling, $W_{11} = W_{12}$, this result coincides with that found previously [14, 15]. It follows from Eq. (7) that the contribution of $N$ identical QEs to the scattering matrix is the same as that of a single QE with the cavity-QE coupling $N^{1/2}\overline{V}_1$ [30, 37]. For negligible interactions with QEs, $\overline{V}_1 = 0$, $S_{12}$ in Eq. (7) coincides with that defined by Eq. (1). For relatively large $N^{1/2}\overline{V}_1$, the transmission amplitude has two resonances separated by Rabi frequency approximately equal to $2N^{1/2}\overline{V}_1$. Remarkably, Eq. (7) shows that transmission amplitude $|S_{12}|$ approaches unity at these resonances if, similar to Eq. (1), the cavity waveguide couplings are equal, $W_{11} = W_{12} = W$, the internal losses are small compared to the losses due to the coupling to waveguides, $\Gamma_1^{(c)}, \Gamma_e^{(a)} \ll 4\pi W^2$, and

$$\Gamma_e^{(a)} \ll \frac{N\overline{V}_1^2}{\pi W^2}. \qquad (8)$$

This equation can be satisfied for large $N$ even if the average cavity-QE coupling $\overline{V}_1$ is small. Since the process of interaction of a photon with QEs and microcavity is inelastic and dissipative, the possibility to reach the unity value of the transmission amplitude of a photon by increasing the number of QEs is not obvious. Figure 5 shows the behavior of transmission probability $P = |S_{12}|^2$ as a function of dimensionless frequency deviation $(\omega - \omega_1^{(c)})/\pi W^2$ and cumulative coupling $N^{1/2}\overline{V}_1/\pi W^2$ for symmetric waveguide coupling $W_{11} = W_{12} = W$, and negligible cavity and QE losses. In this figure, the resonances of transmission amplitude are equal to unity at frequencies which are separated by Rabi frequency. The width of these resonances $\sim \pi W^2$ is independent of the number of QEs, while their separation grows proportionally to $N^{1/2}\overline{V}_1$.

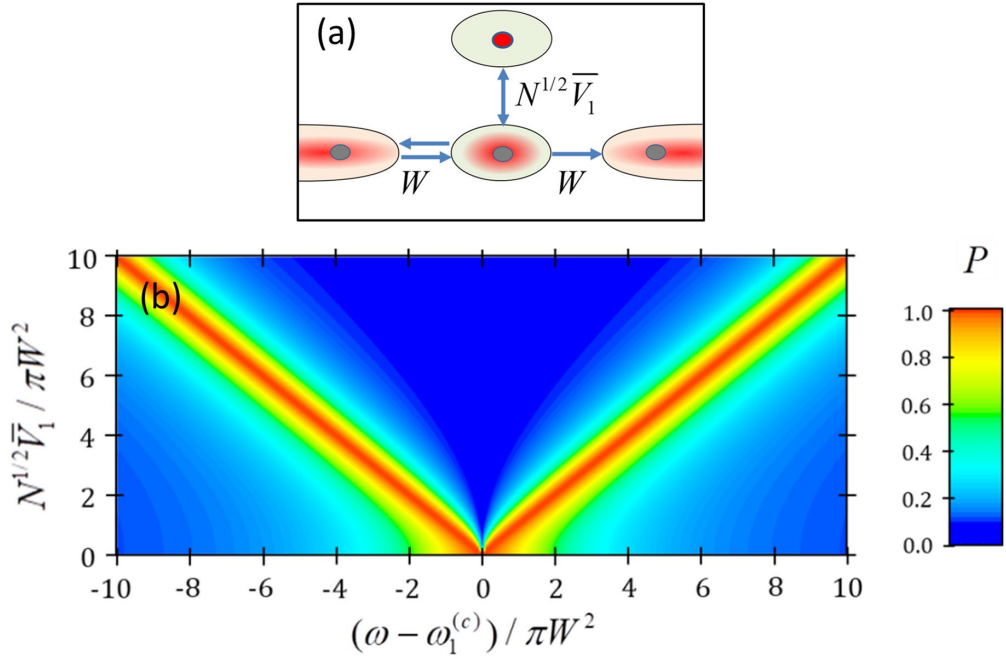

**Fig. 5.** (a) The reduced structure of collective states from Fig. 4(d) with symmetric coupling to waveguides, $W_{11} = W_{12} = W$. (b) Spectrogram of transmission probability $P = |S_{12}|^2$ as a function of dimensionless frequency deviation $(\omega - \omega_1^{(c)}) / \pi W^2$ and cumulative cavity-QE coupling $N^{1/2}\overline{V}_1 / \pi W^2$ for symmetric structure shown in (a) and negligible cavity and QE losses.

The Mahaux-Weidenmüller theory, which was presented in Sec. II in application to systems with semi-infinite waveguides (Fig. 3), can be generalized to the case of infinite waveguides. For example, Fig. 4(e) illustrates an infinite waveguide where an input single photon directly interacts with two-level QEs. To determine the S-matrix of this problem, we approximate the infinite waveguide by two semi-infinite waveguides with a cavity similar to those shown in Fig. 4(b) and set $W_{11} = W_{12} = W$. In the limit of infinite waveguide, coupling $W$ is large, $\pi W^2 >> |\Omega - \Omega_1^{(c)}|$. Then Eq. (7) can be rewritten as

$$S_{11} = \frac{2i\pi W_{eff}^2}{\Omega - \Omega^{(a)} + 2i\pi W_{eff}^2}, \quad S_{12} = 1 - S_{11}, \quad W_{eff}^2 = \frac{N\overline{V}_1^2}{4\pi^2 W^2},, \tag{9}$$

where $W_{eff}$ is the effective coupling parameter between the waveguide and QEs. Remarkably, in this approximation, the reflection amplitude $S_{11}$ coincides with the Breit-Wigner formula, Eq. (1) for symmetric waveguide-cavity coupling, $W_{11} = W_{22} = W_{eff}$. For the particular case of a single QE, $N = 1$, this result coincides with that found in [14]. At exact resonance, $\omega = \omega_e^{(a)} - \omega_g^{(a)}$, we have $S_{11} = (1 + \pi W^2 \Gamma_e^{(a)} / N \overline{V_1}^2)^{-1}$. From this equation, we find that, again, under the condition of Eq. (8), which can be satisfied for sufficiently large number of QEs, the amplitudes $S_{11} \to 1$ and $S_{12} \to 0$. Thus, similar to the previous example, the cumulative action of $N$ QEs distributed in a waveguide can lead to the complete reflection of a single photon.

## IV. DOUBLE EIGENFREQUENCY OPTICAL MICROCAVITY COUPLED TO $N$ THREE-LEVEL QUANTUM EMITTERS

It is well known [5] and immediately follows from the Breit-Wigner formula, Eq. (1), that the resonant transmission amplitude reaches unity at exact resonance $\Omega = \text{Re}\,\Omega_1^{(0)}$ for the symmetric structure with relatively small internal losses $\text{Im}\,\Omega_1^{(0)} << W_{11}^2 = W_{12}^2$. Similarly, as discussed in the previous section, Sec. III, the resonant interaction with a single QE can reflect a photon propagating along the infinite waveguide with the amplitude approaching unity [14]. It is crucial that under the condition of Eq. (8) the latter effect can be enhanced by the cumulative action of QEs. It is of great interest to find out if, similarly, the cumulative action of QEs can support the *inelastic* transition of a single photon to the amplitude close to unity. In the system considered in Sec. III, a photon state in the optical cavity is set in resonance with transitions between two levels of QEs. Propagation of a photon through such system cannot change its frequency due to the energy conservation. In order to enable the frequency conversion of a single photon, it is necessary to consider its interaction with QEs having more than two resonant levels.

As noted in the Introduction, the Breit-Wigner formula, Eq. (1), can describe the inelastic propagation of a particle. Then, the input state should be considered as a collective state of this particle and the environment, which forms a closed system together with this particle. As an example, Eq. (1) can describe the inelastic resonant transmission of a single photon interacting with a three-level QE in an optical waveguide [38-40]. In this case, the collective states $\left| \Lambda_{1,\Omega} \right\rangle$, $\left| \Lambda_{2,\Omega} \right\rangle$ and localized state $\left| \Psi_1 \right\rangle$ are defined as follows: $\left| \Lambda_{1,\Omega} \right\rangle$ is a state of a photon with initial energy $\hbar \omega$ and an atom in the initial state $|g_1\rangle$ with energy $\hbar\omega_{g_1}^{(a)}$; $|\Psi_1\rangle$ is the state of the QE in the excited state $|e\rangle$ with energy $\hbar\omega_e^{(a)}$ which has acquired this photon; $|\Lambda_{2,\Omega}\rangle$ is the state of the output photon with energy $\hbar(\omega + \omega_{g_1}^{(a)} - \omega_{g_2}^{(a)})$ emitted by the QE which is transmitted to the final state $|g_2\rangle$ with energy $\hbar\omega_{g_2}^{(a)}$.

However, the realization of the described system, which performs the inelastic transition of a single photon with the amplitude close to unity, is challenging. In fact, while the input photon frequency $\omega$ can be set in resonance with frequency $\omega_e^{(a)} - \omega_{g_1}^{(a)}$ of QE transition $|g_1\rangle \rightarrow |e\rangle$, the QE transition $|e\rangle \rightarrow |g_2\rangle$ with converted frequency $\omega_e^{(a)} - \omega_{g_2}^{(a)}$ is not resonant. Therefore, transitions from the excited state $|e\rangle$ to states with frequencies other than $\omega_{g_1}^{(a)}$ and $\omega_{g_2}^{(a)}$ cannot be ignored. Furthermore, while driving with an external field [41, 42] can solve the problem, the required symmetry (quantum impedance matching) condition $W_{11} = W_{12}$ is hard to achieve for the undriven QED structures suggested to date [38-40] unless $\omega_{g_1}^{(a)} = \omega_{g_2}^{(a)}$. The question remains of whether, in the absence of an external driving field, the complete conversion of the single-photon frequency is indeed feasible.

In this section, we describe a simplest realistic system of an optical cavity and QEs, which exhibits the complete down-conversion of the single photon frequency in the absence of external driving field. In this system, two states of the optical cavity are in resonance with two transitions between three different levels of QEs. In contrast to the resonant structures considered previously [38-40], the frequencies of both QE transitions are now in resonance with the cavity eigenfrequencies. Therefore, neglecting all other possible QE transitions is justified. Unlike the interaction with two-level QEs considered in the previous section, we find that the cumulative effect of QEs in this case is partial only.

We consider a system consisting of an optical microcavity with two states, $|u_1\rangle$ and $|u_2\rangle$, and $N$ three-level QEs inside. The optical states are coupled to two waveguides so that state $|u_m\rangle$ is coupled to waveguide $m$ only. Possible models of such cavities are illustrated in Fig. 6(a). The model illustrated at the top of Fig. 6(a) consists of an optical resonator coupled to the input and output waveguides through weakly transparent mirrors. The one-dimensional configuration of the cavity states shown at the bottom of Fig. 6(a) requires special design of the optical cavity structure, which is described in Appendix B. For the application of our concern, the resonant structures shown in Fig. 6(a) can be fabricated of Fabry-Perot [44-47], photonic crystal [48, 49], toroidal [50], bottle [51], and SNAP [52, 53] microresonators.

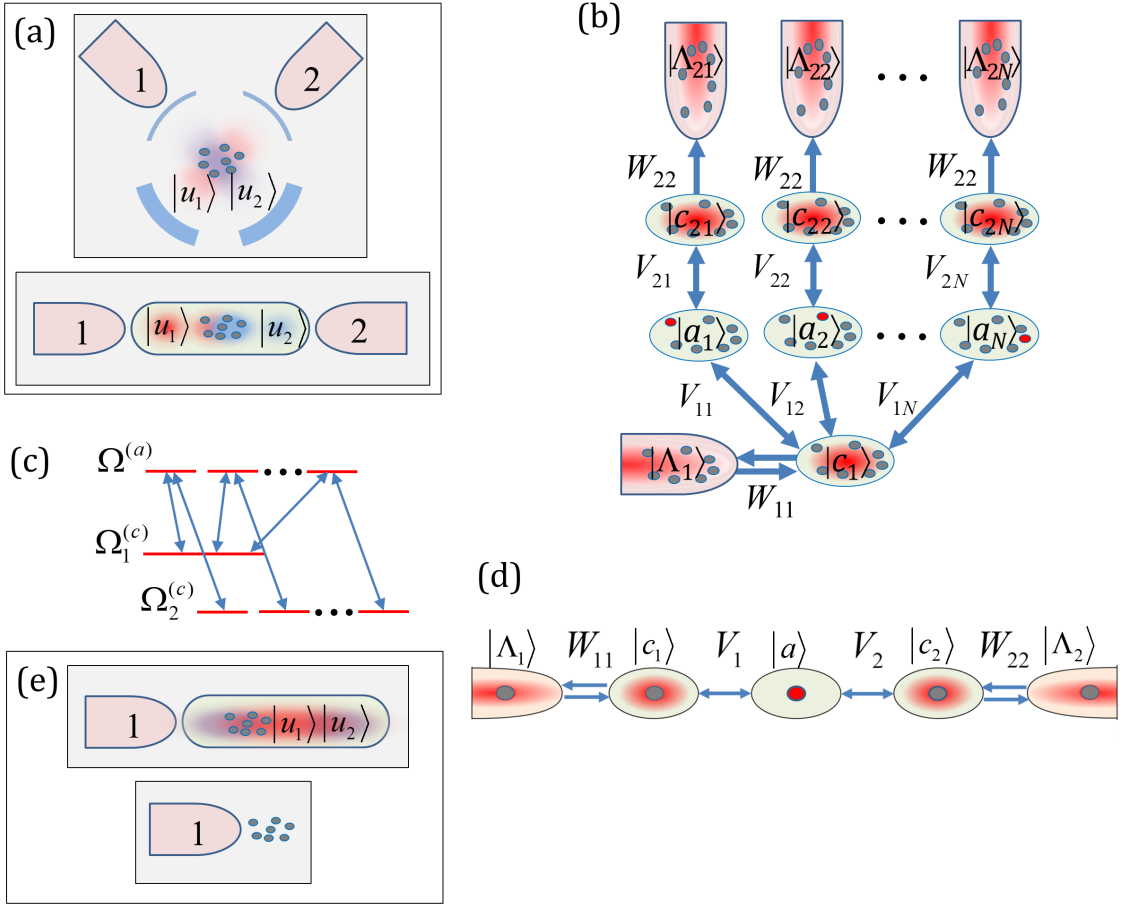

**Fig. 6.** (a) Possible configurations of cavities with two states, $\left|u_1\right\rangle$ and $\left|u_2\right\rangle$, which resonantly interact with transitions of QEs situated in the region where these states overlap in space. (b) Collective localized states and semi-infinite collective waveguide states which describe the propagation of a single photon resonantly coupled to two states of an optical cavity and $N$ three-level QEs positioned inside the cavity. Index $\Omega$ in the notations of $\left|\Lambda_1\right\rangle$ and $\left|\Lambda_{2n}\right\rangle$ is omitted for briefness. (c) Schematics of transition between resonant frequencies of collective states considered, $\Omega_1^{(c)} \leftrightarrow \Omega^{(a)} \leftrightarrow \Omega_2^{(c)}$. (d) For identical QEs with the same positions and, thus, the same coupling parameters $V_{1n} = V_1$ and $V_{2n} = V_2$, the structure of collective states shown in (b) is reduced to the structure shown in this figure. (e) Top: two states of a cavity, $\left|u_1\right\rangle$ and $\left|u_2\right\rangle$, coupled to a single waveguide; bottom: a waveguide coupled to $N$ three-level QEs.

It is assumed that QEs are identical and have two ground states $\left|g_m\right\rangle$, $m = 1, 2$, with energies $\hbar\omega_{g_m}^{(a)}$ and one excited state $\left|e\right\rangle$ with complex energy $\hbar(\omega_e^{(a)} - \frac{i}{2}\Gamma_e^{(a)})$ where $\Gamma_e^{(a)}$ determines the state dissipation. At zero temperature, QEs are initially in their ground state with smallest possible energy $\hbar\omega_{g_1}^{(a)} < \hbar\omega_{g_2}^{(a)}$. In this case, only down-conversion of the photon frequency is possible. If QEs are initially prepared in the excited state, then the condition $\hbar\omega_{g_1}^{(a)} > \hbar\omega_{g_2}^{(a)}$ and the up-conversion of the photon frequency is possible. The complex eigenfrequencies of optical cavity states $\hbar(\omega_m^{(c)} - \frac{i}{2}\Gamma_m^{(c)})$, where $\Gamma_m^{(c)}$ determines the internal cavity loss, are assumed to be in resonance with QE transition frequencies, i.e., $\omega_m^{(c)}$ is close to $\omega_e^{(a)} - \omega_{g_m}^{(a)}$. A single photon with energy $\hbar\omega$, which is close to energy $\hbar\omega_1^{(c)}$ of state $|u_1\rangle$, enters the system through waveguide 1. After entering the cavity state $|u_1\rangle$ through waveguide 1, the photon can be resonantly absorbed by one of the QEs into excited state $|e\rangle$ and bounce between these states and cavity states $|u_m\rangle$ prior to exiting into one of the waveguides.

The described system can be presented as a system of $2N+1$ localized collective eigenstates, $|c_1\rangle$, $|c_{21}\rangle, |c_{22}\rangle, ..., |c_{2N}\rangle$ and $|a_1\rangle, |a_2\rangle, ..., |a_N\rangle$, and $N+1$ collective waveguide states $|\Lambda_1\rangle$, $|\Lambda_{21}\rangle, |\Lambda_{22}\rangle, ..., |\Lambda_{2N}\rangle$ [Fig. 6(b)]. For briefness, we omit index $\Omega$ in the notations of $|\Lambda_1\rangle$ and $|\Lambda_{2n}\rangle$. Collective state $|c_1\rangle$ has a photon localized in the cavity state $|u_1\rangle$ with energy $\hbar\omega_1^{(c)}$ and all QEs in the ground state $|g_1\rangle$ with energy $\hbar\omega_{g_1}^{(a)}$. Collective state $|c_{2n}\rangle$ has a photon localized in the cavity state $|u_2\rangle$ with energy $\hbar\omega_1^{(c)}$, QE $n$ in the ground state $|g_2\rangle$ with energy $\hbar\omega_{g_2}^{(a)}$, and all other QEs in the ground state $|g_1\rangle$ with energy $\hbar\omega_{g_1}^{(a)}$. The energies of states $|c_1\rangle$ and $|c_{2m}\rangle$, if assumed uncoupled to waveguides and QEs, are $\hbar\Omega_1^{(c)} = \hbar(\omega_1^{(c)} + N\omega_{g_1}^{(a)} - \frac{i}{2}\Gamma_1^{(c)})$ and $\hbar\Omega_2^{(c)} = \hbar[\omega_2^{(c)} + \omega_{g_2}^{(a)} + (N-1)\omega_{g_1}^{(a)} - \frac{i}{2}\Gamma_2^{(c)}]$ [Fig. 6(c)]. Collective state $|a_n\rangle$ has no photon in the cavity, a single QE with number $n$ in the excited state $|e\rangle$ and all other QEs in their ground states $|g_1\rangle$. All eigenenergies of states $|a_n\rangle$ have the same value $\hbar\Omega^{(a)} = \hbar[\omega_e^{(a)} + (N-1)\omega_{g_1}^{(a)} - \frac{i}{2}\Gamma_e^{(a)}]$. The waveguide states $|\Lambda_1\rangle$ and $|\Lambda_{2n}\rangle$ have equal total energies $\hbar\Omega = \hbar[\omega + N\omega_{g_1}^{(a)}]$.

In the system described, a photon can either exit from cavity state $|u_1\rangle$ through waveguide 1 with the amplitude $S_{1,1}$ and conserve its original frequency $\omega$, or propagate through QE $n$ and cavity state $|u_2\rangle$ and exit through waveguide 2 with the amplitude $S_{1,2n}$ and frequency conversion $\omega \rightarrow \omega + \omega_{g_1}^{(a)} - \omega_{g_2}^{(a)}$. Direct application of Eq. (5) to the system shown in Fig. 6(b) yields (see Appendix A)

$$S_{1,2m} = \frac{2i\pi W_{11}W_{22}V_{1m}V_{2m}}{\left(\Delta\omega_2\Delta\omega_a - V_{2m}^2\right)\left(\Delta\omega_1 - \Delta\omega_2\sum_{n=1}^{N}\dfrac{V_{1n}^2}{\Delta\omega_2\Delta\omega_a - V_{2n}^2}\right)}, \qquad (10)$$

$$\begin{aligned}
&\Delta\omega_1 = \omega - \omega_1^{(c)} + i(\pi W_{11}^2 + \tfrac{1}{2}\Gamma_1^{(c)}),\\
&\Delta\omega_2 = \omega - \omega_2^{(c)} + \omega_{g_1}^{(a)} - \omega_{g_2}^{(a)} + i(\pi W_{22}^2 + \tfrac{1}{2}\Gamma_2^{(c)}),\\
&\Delta\omega_a = \omega - \omega_e^{(a)} + \omega_{g_1}^{(a)} + \tfrac{i}{2}\Gamma_e^{(a)},
\end{aligned} \qquad (10a)$$

where $V_{1n}$ is the coupling between states $|c_1\rangle$ and $|a_n\rangle$, $V_{2n}$ is the coupling between states $|c_{2n}\rangle$ and $|a_{2n}\rangle$, $W_{11}$ is the coupling between states $|\Lambda_1\rangle$ and $|c_1\rangle$, and $W_{22}$ is the coupling between states $|\Lambda_{2n}\rangle$ and $|c_{2n}\rangle$ [Fig. 6(b)]. All couplings between states $|\Lambda_{2n}\rangle$ and $|c_{2n}\rangle$ are equal because they are determined by the coupling between photon state $|u_2\rangle$ and waveguide 2. The full inelastic transmission probability is determined from Eq. (10) as

$$P = \sum_{n=1}^{N}|S_{1,2n}|^2 \quad. \qquad (11)$$

## V.  COMPLETE FREQUENCY DOWN-CONVERSION IN THE ABSENCE OF LOSSES

We assume that couplings to QEs are independent of their number as, e.g., for QEs situated near an antinode of the optical cavity (see, e.g., [33-35]) and set $V_{1n} = V_1$ and $V_{2n} = V_2$. Then Eqs. (10) and (11) yield:

$$P = \frac{4\pi^2 N W_{11}^2 W_{22}^2 V_1^2 V_2^2}{\left|\Delta\omega_1\left(\Delta\omega_2\Delta\omega_a - V_2^2\right) - N\Delta\omega_2 V_1^2\right|^2}. \qquad (12)$$

Rescaling $N^{1/2}V_1 \to V_1$ transfers this equation to the case of a single QE, $N = 1$ and reduces the problem to the well-known problem of resonant propagation through three successively coupled collective localized states illustrated in Fig. 6(d) (see e.g., [26]). Remarkably, Eq. (12) shows that coupling $V_1$ can be effectively enhanced by the increasing the number of QEs $N$. This result manifests the partial cumulative action of QEs when only $V_1$ rather than both $V_1$ and $V_2$ is enhanced. The latter fact has a simple physical explanation. While the amplitude of transmission of a photon from the cavity state $|u_1\rangle$ into one of the excited states of QE increases with the number of QEs $N$, the amplitude of transmission from this state of QE into cavity state $|u_2\rangle$ does not depend on the number of QEs. This situation is different from the cumulative action of two-level QEs considered in Sec. III where a photon emitted from an excited QE can return to the same collective state and be acquired by other QEs.

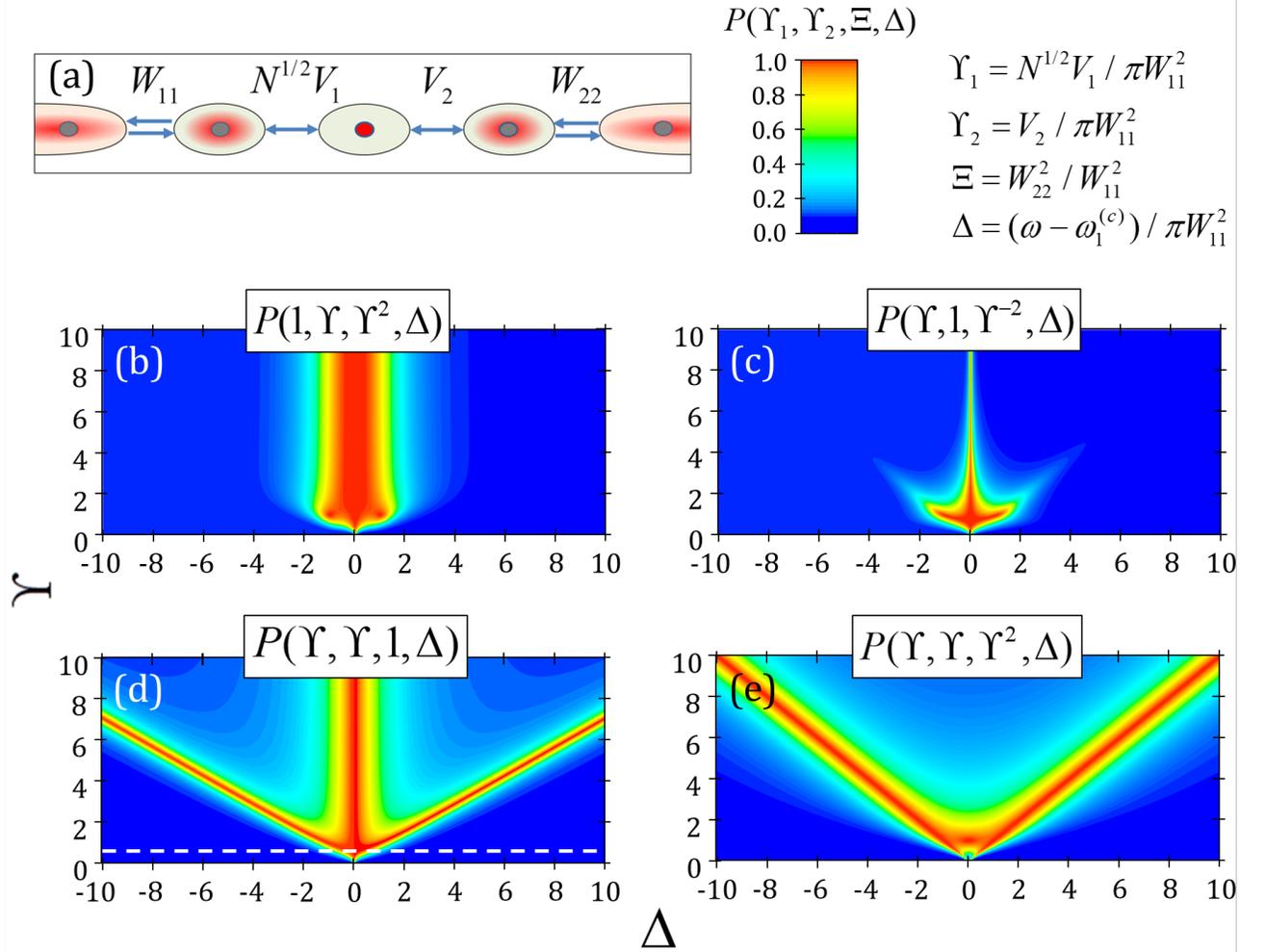

**Fig. 7.** (a) A system of coupled collective states of a two-level cavity and a single three-level QE, which is equivalent to the system of $N$ identical QEs with equal couplings to the cavity states, $V_{1n} = V_1$ and $V_{2n} = V_2$. (b–e) Spectrograms of inelastic transmission probability $P(\Upsilon_1, \Upsilon_2, \Xi, \Delta)$ for particular relations between $\Upsilon_1$, $\Upsilon_2$, and $\Xi$ indicated on the figures. Horizontal white dashed line in (d) corresponds to ultraflat behavior of $P$ described by Eq. (18).

It follows from Eqs. (12) and (10a) that the cavity losses and dissipation of QEs can be ignored if

$$\Gamma_1^{(c)} << 2\pi W_{11}^2, \quad \Gamma_2^{(c)} << 2\pi W_{22}^2 \qquad (13a)$$

and

$$\Gamma_e^{(a)} << 2\pi(W_{11}^2 + W_{22}^2), \quad \Gamma_e^{(a)} << \frac{2(NV_1^2 + V_2^2 + \pi^2 W_{11}^2 W_{22}^2)}{\pi(W_{11}^2 + W_{22}^2)}. \qquad (13b)$$

Equation (13a) is similar to the condition of complete resonant transparency of an empty optical cavity discussed in the Introduction, while Eq. (13b) requires that the dissipation of QEs was relatively small compared to the cavity-QE and/or cavity-waveguide couplings. This means that the dissipation time of QEs should be relatively large compared to the characteristic time of inelastic transition.

Neglecting losses, $\Gamma_e^{(a)} = \Gamma_m^{(c)} = 0$, we can find the general condition when the inelastic transmission probability $P$ determined by Eq. (12) achieves unity. Specifically, the criterion of the equality $P = 1$ derived in Appendix C is given by two simultaneous equations:

$$W_2^2 V_2^2 \left[ \left( \omega - \omega_1^{(c)} \right)^2 + \pi^2 W_1^4 \right] = N W_1^2 V_1^2 \left[ \left( \omega - \omega_2^{(c)} + \omega_{g_1}^{(a)} - \omega_{g_2}^{(a)} \right)^2 + \pi^2 W_2^4 \right]$$

$$W_1^2 \left( \omega - \omega_e^{(a)} + \omega_{g_1}^{(a)} \right) \left[ \left( \omega - \omega_2^{(c)} + \omega_{g_1}^{(a)} - \omega_{g_2}^{(a)} \right)^2 + \pi^2 W_2^4 \right]$$
$$= V_2^2 \left[ W_1^2 \left( \omega - \omega_2^{(c)} + \omega_{g_1}^{(a)} - \omega_{g_2}^{(a)} \right) + W_2^2 \left( \omega - \omega_1^{(c)} \right) \right] \qquad (14)$$

Here we limit our consideration by a special case when the cavity eigenfrequencies are tuned to the exact resonance with frequencies of the QE transitions, $\omega_1^{(c)} + \omega_{g_1}^{(a)} = \omega_2^{(c)} + \omega_{g_2}^{(a)} = \omega_e^{(a)}$. Then the inelastic transmission probability $P$ determined by Eq. (12) can be expressed through four dimensionless parameters: relative cavity-QE couplings $\Upsilon_1 = N^{1/2} V_1 / \pi W_{11}^2$ and $\Upsilon_2 = V_2 / \pi W_{11}^2$, relative cavity-waveguide coupling $\Xi = W_{22}^2 / W_{11}^2$, and relative frequency deviation $\Delta = (\omega - \omega_1^{(c)}) / \pi W_{11}^2$ as:

$$P = P(\Upsilon_1, \Upsilon_2, \Xi, \Delta) = \frac{4\Xi \Upsilon_1^2 \Upsilon_2^2}{\left[ \Delta^3 - \left( \Xi + \Upsilon_1^2 + \Upsilon_2^2 \right) \Delta \right]^2 + \left[ \left( 1 + \Xi \right) \Delta^2 - \Upsilon_2^2 - \Xi \Upsilon_1^2 \right]^2}. \qquad (15)$$

For this particular case, Eq. (14) is reduced to

$$\Xi\left(1+\Upsilon_2^2\right) = \Xi^2 + \Upsilon_1^2,$$
$$\Delta^2 = \Upsilon_2^2\left(1+\Xi\right) - \Xi^2,$$

(16)

for $\Delta \neq 0$ and

$$\Upsilon_2^2 = \Xi\Upsilon_1^2$$

(17)

for $\Delta = 0$. Figures 7(b)-7(e) show characteristic spectrograms of inelastic transmission probability $P(\Upsilon_1, \Upsilon_2, \Xi, \Delta)$ corresponding to particular relations between $\Upsilon_m$ and $\Xi$ when $P$ can achieve unity. It is seen that for the exact resonant structure the spectrograms are symmetric with respect to $\Delta$ (see Appendix A). Generally, satisfaction of Eq. (17) is sufficient for $P$ to achieve unity at $\Delta = 0$ as illustrated in Figs. 7(b) and (c) which show $P(1, \Upsilon, \Upsilon^2, \Delta)$ and $P(\Upsilon, 1, \Upsilon^{-2}, \Delta)$, respectively. Remarkably, for predetermined cavity-QE couplings $V_1$ and $V_2$, satisfaction of Eq. (17) can be achieved by tuning waveguide-cavity couplings $W_{11}$ and $W_{22}$. As follows from Eqs. (16) and (17), for the symmetric coupling to waveguides when $\Xi = 1$, the value of $P$ achieves unity only if $\Upsilon_1 = \Upsilon_2 = \Upsilon$. Specifically, $P = 1$ along the vertical line $\Delta = 0$ and along two symmetrically positioned lines $\Upsilon = \pm 2^{-1/2}(\Delta^2 + 1)^{1/2}$ as shown in the spectrogram of $P(\Upsilon, \Upsilon, 1, \Delta)$ in Fig. 7(d). The ultraflat behavior of inelastic transmission probability approaching unity is achieved along the white horizontal line of Fig. 7(d) which corresponds to

$$P(0.5, 0.5, 1, \Delta) = \frac{1}{1+\Delta^6}$$

(18)

known as the Butterworth filter profile in the theory of signal processing (see, e.g., [28]). Finally, the sufficient condition $\Upsilon_1^2 = \Upsilon_2^2 = \Xi$ for the unity probability $P$ following from Eqs. (16) is illustrated in Fig. 7(e) which shows $P(\Upsilon, \Upsilon, \Upsilon^2, \Delta)$. In this case, Eq. (17) is not satisfied and $P = 1$ along the lines $\Upsilon = \pm\Delta$ only. Experimental realization of complete frequency down-conversion of a single photon illustrated in Fig. 7 is possible if the cavity loss and QE dissipation are small enough. Specifically, it is required that the conditions of Eqs. (13a) and (13b) for the system losses are satisfied in the regions of surface plots in Figs. 7(b)-7(e) where $P$ approaches unity. For example, assuming that the introduced dimensionless parameters of the system have the same order of magnitude equal to unity, $\Upsilon_1 \sim \Upsilon_2 \sim \Xi \sim 1$, these conditions simply require that $\Gamma_e^{(a)}, \Gamma_m^{(c)} << \pi W_{11}^2$. Remarkably, the condition $\Upsilon_1 \sim 1$ can be achieved for sufficiently large number of QEs $N$ even if $V_1 << V_2$.

Of special interest is the situation when only a single waveguide is coupled to an optical cavity as illustrated in the top of Fig. 6(e). The model of collective states shown in Fig. 6(b) is valid in this case as well, though now the inelastic transmission amplitude calculated in Sec. V corresponds to the inelastic reflection amplitude back into the same waveguide. Assuming that the waveguide-cavity coupling is a weak function of wavelength, we set $W_{11} = W_{22}$, i.e., $\Xi = 1$. Then we find from Eqs. (15) and (16) that the inelastic reflection amplitude $P$ can be equal to unity only if $\Upsilon_1 = \Upsilon_2 = \Upsilon$. Thus, similar to the case of two waveguide, $P$ is equal to unity along the line $\Delta = 0$ and two symmetric lines $\Upsilon = \pm 2^{-1/2}(\Delta^2 + 1)^{1/2}$ shown in Fig. 7(d).

Let us now investigate the direct waveguide-QE resonant coupling. Without loss of generality, we consider the case of a single input-output waveguide illustrated in the bottom of Fig. 6(e). Similar to the derivation of Eq. (9), we assume large and equal waveguide-cavity couplings $W_{11} = W_{12} = W$, so that Eq. (12) is reduced to

$$P = \frac{4 W_{eff,1}^2 W_{eff,2}^2}{\Delta \omega_a^2 + \left(W_{eff,1}^2 + W_{eff,2}^2\right)^2}, \quad W_{eff,1}^2 = \frac{N V_1^2}{\pi W^2}, \quad W_{eff,1}^2 = \frac{V_2^2}{\pi W^2}. \tag{19}$$

where $W_{eff,p}^2$ are the effective waveguide-QE couplings similar to that introduced in Eq. (9). Again, as discussed above, in contrast to coupling to two-level atoms considered in Sec. III, the cumulative action of atoms can enhance the coupling $V_1$ rather than both $V_1$ and $V_2$.

In a more general case, QEs are situated in different positions inside the optical cavity. In the absence of losses, $\Gamma_1^{(c)} = \Gamma_2^{(c)} = \Gamma_e^{(a)} = 0$, and for the exact resonant condition, $\Delta \omega_1 = \Delta \omega_2 = \Delta \omega_a = 0$, Eqs. (10) and (11) yield:

$$P = P_0 = \frac{4\pi^2 W_{11}^2 W_{22}^2 \Sigma_V}{\left(W_{11}^2 + W_{22}^2 \Sigma_V\right)^2}, \quad \Sigma_V = \sum_{n=1}^{N} \frac{V_{1n}^2}{V_{2n}^2}. \tag{20}$$

This equation shows that the complete frequency conversion takes place if

$$\frac{W_{11}^2}{W_{22}^2} = \sum_{n=1}^{N} \frac{V_{1n}^2}{V_{2n}^2} \qquad (21)$$

Remarkably, provided that waveguide-cavity couplings $W_{11}$ and $W_{22}$ satisfy this equation, random positions of QEs (leading to different $V_{mn}$ for different $n$) will not reduce the effect of complete frequency conversion. If the field distributions of cavity modes $|u_1\rangle$ and $|u_2\rangle$ are proportional to each other in the region where QEs are situated so that $u_1(\mathbf{r}) = \beta u_2(\mathbf{r})$, then the ratio $V_{1n}/V_{2n} = \beta$ does not depend on $n$ and Eq. (20) is reduced to $W_{11}^2 = N\beta^2 W_{22}^2$. The top of Fig. 6(a) shows the configuration of two localized optical states and QEs positioned in the region where these states overlap. Each of these states is coupled to only one input-output waveguide. It is apparent that the mirrors of this cavity can be designed so that the relation $u_1(\mathbf{r}) = \beta u_2(\mathbf{r})$ is accurately satisfied near the antinodes of these states. The design of an alternative one-dimensional configuration of optical states shown in the bottom of Fig. 6(a) is less obvious and considered in Appendix B. We suggest that the experimental realization of such structure is possible, for example, employing quantum dots in coupled photonic crystal cavities. In Ref. [49], a single quantum dot with dissipation rate $\Gamma_e^{(a)} \sim 10$ MHz and strong coupling $V \sim 20$ GHz to a single-cavity mode was demonstrated. For quantum dots with similar characteristic parameters $\Gamma_e^{(a)} << V$ and photonic crystal cavities with losses $\Gamma_j^{(c)} << V$ corresponding to the intrinsic $Q$ factor exceeding $10^5$, the conditions of lossless transmission, Eqs. (13a) and (13b), can be satisfied. Then, in the simplest case of a single quantum dot ($N = 1$) with three resonant levels and quantum dot – cavity coupling parameters with ratio $V_1/V_2 = \beta$, the waveguide-cavity coupling can be adjusted as described above.

The analogy of the frequency conversion scheme considered here and the SPRINT scheme (see Refs. [43, 54] and references therein) should be noted. In fact, for the case of one atom, $N = 1$, symmetric coupling, $W_{11} = W_{22} = W$, equal cavity losses, $\Gamma_1^{(c)} = \Gamma_1^{(c)} = \Gamma^{(c)}$, and at the exact resonance, $\omega = \omega_1^{(c)} = \omega_2^{(c)} - \omega_{g_1}^{(a)} + \omega_{g_2}^{(a)} = \omega_e^{(a)}$, Eqs. (10a) and (12) yield:

$$P = \frac{4\pi^2 W^4 V_1^2 V_2^2}{\left(\pi W^2 + \frac{1}{2}\Gamma^{(c)}\right)^2 \left(\left(\pi W^2 + \frac{1}{2}\Gamma^{(c)}\right)\Gamma_e^{(a)} + V_1^2 + V_2^2\right)^2}. \qquad (22)$$

This result is identical to Eq. (7b) of Ref. [54] derived for the degenerated states of a three-level atom, $\omega_{g_1}^{(a)} = \omega_{g_2}^{(a)}$, and an optical cavity, $\omega_1^{(c)} = \omega_2^{(c)}$, i.e., for the case of no frequency conversion.

## VI. DISSCUSSION

The Mahaux-Weidenmüller formalism [18-20], which was developed here for the resonant propagation of a single photon in cavity QED, allows one to consider systems consisting of a large number of quantum particles and collective excitations distributed inside microcavities without the application of the second quantization method. In this formalism, the Hamiltonian of the system is presented as a linear combination of ket-bra products of its eigenfunctions [Eq. (4)]. In the simplest case, these eigenfunctions can be the states of a single physical particle. In more general cases considered in this paper, they are the collective states of several particles or excitations, whose total energy $\hbar\Omega$ is conserved in the process of propagation.

The resonant approximation for the input-output problems consists in selecting the states of the system whose energy is close to the energy $\hbar\Omega$ of the input collective state and ignoring all other nonresonant states. It is often favorable to perform this selection directly in the original Hamiltonian of the system. An example of replacing the Hamiltonian of identical photons and a two-level QE in an optical cavity having infinite number of eigenstates by just two resonant collective states of a single photon and QE is given in the Introduction. Developing this approach further, we considered the input-output problems for a single photon resonantly propagating through optical cavities with two- and three-level QEs inside. The eigenstates of these systems were selected in the form of a collective state of a photon and $N$ QEs interacting with it.

The energy $\hbar\Omega$ of the input collective state can be redistributed within the components of this state in the process of transmission. Then, a particle (being a part of a collective state) can be transmitted inelastically and, as a result, resonantly acquire or release (respectively, the environment within the collective state can release or acquire) an energy. It is worth noting that the exchange of energy between the propagating particle (in our case – a single photon) and environment is analogous to the exchange of energy between degrees of freedom of a wave resonantly propagating through a localized state in three-dimensional space, whose transmission amplitude can be determined by the Breit-Wigner formula, Eq. (1) [55, 56]. While the total energy of the input wave is conserved, the energy of its transverse degrees of freedom can be transferred to or acquired by its longitudinal degree of freedom. Another example analogous to the inelastic transition of a photon interacting with QEs considered in this paper is the inelastic transition of an electron resonantly interacting with electromagnetic field in a quantum well structure. In fact, it was shown in Ref. [57] that the amplitude of such transition can approach unity. Remarkably, the expression for the transmission amplitude derived in Ref. [57] through cumbersome calculations directly follows from the Mahaux-Weidenmüller formula.

In Sec. III of this paper, we demonstrated the strength of the suggested approach considering the classical problem of resonant propagation of a single photon interacting with two-level QEs in an optical cavity [14, 15, 29-36]. The direct application of the Mahaux-Weidenmüller formula to this problem allowed us to arrive at the general analytic expressions for the transmission and reflection amplitudes of a single photon, which coincided with previous results in particular cases. The important effect manifested by the derived expressions is the cumulative action of QEs. We have shown that the probability of resonant transmission of a single photon can approach unity if the number of QEs interacting with it is sufficiently large. The cumulative action of atoms was described in several experiments [31-35]. However, not much has been published regarding the solution of the general input-output problem for a single photon resonantly interacting with two-level QEs [Eq. (6)]. While the interaction of a photon and a QE inside the cavity is inelastic in this problem, the input and output photons have the same energy, i.e., the total process results in the elastic transmission of the photon.

The described cumulative action of QEs, which enhance the resonant *elastic* transmission probability of a single photon, led us to the question of whether a similar effect can increase the probability of *inelastic* transmission of a single photon up to the value close to unity and, thus, ensure its complete frequency conversion. The principal barriers on the way to demonstrating the complete resonant conversion of the single photon frequency in microscopic cavity QED systems arise from the difficulty to achieve the sufficient resonant enhancement of interactions between the photon and QEs and, in addition, to arrive at the condition of symmetric coupling [e.g., the equality of $W_{11}$ and $W_{12}$ in Eq. (1)]. While these barriers can be overcome by application of the external driving field [41], the realistic undriven microscopic structure where all selected transitions of QEs are in resonance with optical cavity eigenstates and, thus, maximally enhanced to enable the complete conversion of the single photon frequency has not been sufficiently investigated.

In Secs. IV and V of this paper, we described the simplest resonant cavity QED system, which enables the complete down-conversion of a single photon frequency in the absence of an external driving field. This system consists of an optical cavity having two states, which resonantly couple to two electronic transitions of QEs positioned inside the cavity. Using the Mahaux-Weidenmüller formalism, the general expression for the inelastic resonant transmission amplitudes of a single photon through the cavity with $N$ QEs distributed inside is derived. For identical QEs having equal couplings $V_1$ to the input cavity state $|u_1\rangle$ and equal couplings $V_2$ to the output cavity state $|u_2\rangle$, the action of $N$ QEs is reduced to the action of a single QE with effective couplings $N^{1/2}V_1$ and $V_2$, which are responsible, respectively, for the absorption and emission of a single photon by QEs. Thus, the cumulative action of QEs enhances the adsorption of a photon by QEs rather than its emission. It is shown that, if the cavity losses and QE dissipation are small enough, the complete frequency conversion can be achieved in the system with appropriately designed waveguide-cavity couplings $W_{11}$ and $W_{22}$ and cavity eigenfrequencies $\omega_1^{(c)}$ and $\omega_2^{(c)}$.

Overall, the Mahaux-Weidenmüller formalism allowed us to solve complex input-output problems in a much simpler way as compared to that based on the second quantization formalism for photons. It is of a great interest to develop a similar approach for the input-output cavity QED problems with more than one photon and more complex configurations of optical microcavities and quantum excitations. It is believed that the simplicity and generality of the developed approach suggest a practical way to identify and describe new phenomena in cavity quantum electrodynamics.

The author acknowledges the Royal Society Wolfson Research Merit Award (WM130110) and funding from the Horizon 2020 Framework Programme (H2020) (H2020-EU.1.3.3, 691011), Engineering and Physical Sciences Research Council (EPSRC) (EP/P006183/1), and US Army Research Laboratory (W911NF-17-2-0048). Discussions with M. Brodsky, A. Fotiadi, V. Malinovsky, A. Sukhorukov, and I. Yurkevich are greatly appreciated. The author greatly appreciates a recent discussion with Arno Rauschenbeutel, where he noted the analogy of the frequency conversion scheme considered here and the SPRINT scheme (see the discussion in the end of Section V).

**APPENDIX A**

Our calculations of the S-matrix determined by the Mahaux-Weidenmüller formula, Eq. (5), are based on the expression for the inverse of a partitioned matrix [57]:

$$\begin{pmatrix} \mathbf{A} & \mathbf{B} \\ \mathbf{C} & \mathbf{D} \end{pmatrix}^{-1} = \begin{pmatrix} \mathbf{F} & -\mathbf{FBD}^{-1} \\ -\mathbf{D}^{-1}\mathbf{CF} & \mathbf{D}^{-1} + \mathbf{D}^{-1}\mathbf{CFBD}^{-1} \end{pmatrix}, \quad \mathbf{F} = \left( \mathbf{A} - \mathbf{BD}^{-1}\mathbf{C} \right)^{-1} \tag{A1}$$

The S-matrix of the resonant propagation of a single photon through a system of an optical cavity and $N$ two-level QEs, which is considered in Sec. IV, can be found by setting in Eq. (5)

$$\mathbf{\Omega I} - \mathbf{\Omega}^{(0)} - \mathbf{V} + i\pi \mathbf{W}^\dagger \mathbf{W} = \begin{pmatrix} \mathbf{A} & \mathbf{B} \\ \mathbf{C} & \mathbf{D} \end{pmatrix} \tag{A2}$$

with

$$\mathbf{A} = \left( \Omega - \Omega_1^{(c)} + i\pi(W_{11}^2 + W_{12}^2) \right),$$

$$\mathbf{B} = \begin{pmatrix} V_{11} & V_{12} & \dots & V_{1N} \end{pmatrix},$$

$$\mathbf{C} = \mathbf{B}^T,$$

$$\mathbf{D} = \begin{pmatrix} \Omega - \Omega_1^{(a)} & 0 & \dots & 0 \\ 0 & \Omega - \Omega_2^{(a)} & \dots & 0 \\ \dots & \dots & \dots & \dots \\ 0 & 0 & \dots & \Omega - \Omega_N^{(a)} \end{pmatrix},$$

(A3)

where $\Omega_1^{(c)} = \omega_1^{(c)} + \sum_{n=1}^{N} \omega_{g_n}^{(a)} - \frac{i}{2}\Gamma_1^{(c)}$, $\Omega_n^{(a)} = \omega_{e_n}^{(a)} - \omega_{g_n}^{(a)} + \sum_{m=1}^{N} \omega_{g_m}^{(a)} - \frac{i}{2}\Gamma_{e_n}^{(a)}$, and $\hbar\omega_{g_n}^{(a)}$ and $\hbar\omega_{e_n}^{(a)} - \frac{i}{2}\Gamma_{e_n}^{(a)}$ are the eigenfrequencies of two-level QE $n$. After substitution of Eqs. (A3) into Eq. (A1) and simple calculations, Eq. (5) is reduced to Eq. (6).

The S-matrix of the resonant propagation of a single photon through a system of an optical cavity with two eigenfrequencies and $N$ three-level QEs, which is considered in Sec. IV, can be found from Eq. (5) using Eqs. (A1) and (A2) with submatrices:

$$\mathbf{A} = \begin{pmatrix} \Omega - \Omega_1^{(c)} + i\pi W_{11}^2 & V_{11} & V_{12} & \dots & V_{1N} \\ V_{11} & \Omega - \Omega^{(a)} + \frac{i}{2}\Gamma_e^{(a)} & 0 & \dots & 0 \\ V_{12} & 0 & \Omega - \Omega^{(a)} + \frac{i}{2}\Gamma_e^{(a)} & \dots & 0 \\ \dots & \dots & \dots & \dots & \dots \\ V_{1N} & 0 & 0 & \dots & \Omega - \Omega^{(a)} + \frac{i}{2}\Gamma_e^{(a)} \end{pmatrix},$$

$$\mathbf{B} = \begin{pmatrix} 0 & 0 & 0 & \dots & 0 \\ V_{21} & 0 & 0 & \dots & 0 \\ 0 & V_{22} & 0 & \dots & 0 \\ \dots & \dots & \dots & \dots & \dots \\ 0 & 0 & 0 & \dots & V_{2N} \end{pmatrix}, \quad \mathbf{C} = \mathbf{B}^T,$$

(A4)

$$\mathbf{D} = \begin{pmatrix} \Omega - \Omega_2^{(c)} + i\pi W_{22}^2 & 0 & \dots & 0 \\ 0 & \Omega - \Omega_2^{(c)} + i\pi W_{22}^2 & \dots & 0 \\ \dots & \dots & \dots & \dots \\ 0 & 0 & \dots & \Omega - \Omega_2^{(c)} + i\pi W_{22}^2 \end{pmatrix},$$

where

$$\Omega = \omega + N\omega_{g_1}^{(a)}$$
$$\Omega_1^{(c)} = \omega_1^{(c)} + N\omega_{g_{1n}}^{(a)} - \frac{i}{2}\Gamma_1^{(c)},$$
$$\Omega_2^{(c)} = \omega_1^{(c)} + \omega_{g_2}^{(a)} + (N-1)\omega_{g_1}^{(a)} - \frac{i}{2}\Gamma_2^{(c)},$$
$$\Omega^{(a)} = \omega_e^{(a)} + (N-1)\omega_{g_1}^{(a)} - \frac{i}{2}\Gamma_e^{(a)},$$

(A5)

and $\hbar\omega_{g_1}^{(a)}$, $\hbar\omega_{g_2}^{(a)}$, and $\hbar\omega_e^{(a)} - \frac{i}{2}\Gamma_e^{(a)}$ are the eigenfrequencies of three-level QE $n$. Direct calculations

yield the following expression for matrix $\mathbf{F}$ in Eq. (A1):

$$\mathbf{F} = \left(\mathbf{A} - \mathbf{B}\mathbf{D}^{-1}\mathbf{C}\right)^{-1} =$$

$$= \begin{pmatrix} \Omega - \Omega_1^{(c)} + i\pi W_{11}^2 & V_{11} & V_{12} & ... & V_{1N} \\ V_{11} & \Omega - \Omega^{(a)} + \frac{i}{2}\Gamma_e^{(a)} - Q_1 & 0 & ... & 0 \\ V_{12} & 0 & \Omega - \Omega^{(a)} + \frac{i}{2}\Gamma_e^{(a)} - Q_2 & ... & 0 \\ ... & ... & ... & ... & ... \\ V_{1N} & 0 & 0 & ... & \Omega - \Omega^{(a)} + \frac{i}{2}\Gamma_e^{(a)} - Q_N \end{pmatrix}^{-1}, \quad \text{(A6)}$$

$$Q_n = \frac{V_{2n}^2}{\Omega - \Omega_2^{(c)} + i\pi W_{22}^2}.$$

In order to find the inverse of the matrix in Eq. (A6), we partition it as

$$\mathbf{F}^{-1} = \begin{pmatrix} \mathbf{A}_1 & \mathbf{B}_1 \\ \mathbf{C}_1 & \mathbf{D}_1 \end{pmatrix}$$

(A7)

where

$$\mathbf{A}_1 = \left(\Omega - \Omega_1^{(c)} + i\pi W_{11}^2\right),$$
$$\mathbf{B}_1 = \begin{pmatrix} V_{11} & V_{12} & ... & V_{1N} \end{pmatrix}, \quad \mathbf{C}_1 = \mathbf{B}_1^T,$$
$$\mathbf{D}_1 = \begin{pmatrix} \Omega - \Omega^{(a)} + \frac{i}{2}\Gamma_e^{(a)} - Q_1 & 0 & ... & 0 \\ 0 & \Omega - \Omega^{(a)} + \frac{i}{2}\Gamma_e^{(a)} - Q_2 & ... & 0 \\ ... & ... & ... & ... \\ 0 & 0 & 0 & \Omega - \Omega^{(a)} + \frac{i}{2}\Gamma_e^{(a)} - Q_N \end{pmatrix},$$

(A8)

and, again, apply Eq. (A1). As the result we find:

$$S_{1,2m} = \frac{2i\pi W_{11}W_{22}V_{1m}V_{2m}}{\left(\left(\Omega - \Omega_1^{(c)} + i\pi W_{11}^2\right)\left(\Omega - \Omega^{(a)} + \frac{i}{2}\Gamma_e^{(a)}\right) - V_{2m}^2\right)\left(\Omega - \Omega_1^{(c)} + i\pi W_{11}^2 - \left(\Omega - \Omega_2^{(c)} + i\pi W_{22}^2\right)\Sigma\right)},$$

$$\Sigma = \sum_{n=1}^{N} \frac{V_{1n}^2}{\left(\Omega - \Omega_2^{(c)} + i\pi W_{22}^2\right)\left(\Omega - \Omega^{(a)} + \frac{i}{2}\Gamma_e^{(a)}\right) - V_{2n}^2}.$$

(A9)

After substitution of notations from Eq. (A5), this equation coincides with Eq. (10) of the main text.

If couplings to QEs are independent of their number Eq. (A9) is simplified so that the full inelastic transmission probability $P = \sum_{n=1}^{N} |S_{1,2n}|^2$ coincides with Eq. (12) of the main text.

In the absence of losses, $\Gamma_e^{(a)} = \Gamma_m^{(c)} = 0$, and at the exact resonance, $\omega_1^{(c)} + \omega_{g_1}^{(a)} = \omega_2^{(c)} + \omega_{g_2}^{(a)} = \omega_e^{(a)}$, the equations in (10a) are simplified to

$$\begin{aligned}
\Delta\omega_1 &= \Delta\omega + i\pi W_{11}^2, \\
\Delta\omega_2 &= \Delta\omega + i\pi W_{22}^2, \\
\Delta\omega_a &= \Delta\omega, \\
\Delta\omega &= \omega - \omega_1^{(c)}.
\end{aligned}$$

(A10)

Substitution of Eq. (A10) into Eq. (12) yields:

$$P = \frac{4\pi^2 N W_{11}^2 W_{22}^2 V_1^2 V_2^2}{\left[\Delta\omega^3 - \left(\pi^2 W_{11}^2 W_{22}^2 + NV_1^2 + V_2^2\right)\Delta\omega\right]^2 + \left[\left(\pi W_{11}^2 + \pi W_{22}^2\right)\Delta\omega^2 - \pi W_{11}^2 V_2^2 - \pi W_{22}^2 V_1^2\right]^2}$$

(A11)

As follows from Eq. (A11), the transmission probability is, in general, an asymmetric function of $\Delta\omega$. However, Eq. (A11) shows that it is the symmetric function of $\Delta\omega$ for the lossless system satisfying the exact resonance condition indicated.

**APPENDIX B**

Here we present a model of microcavity illustrated in the bottom of Fig. 6(a). We compose it of three weakly coupled short-range cavities illustrated in Figs. 8(a) and 8(b). The optical states localized in this cavity are defined by the model wave equation:

$$\frac{d^2u}{dx^2} - \left[ \kappa^2 + \alpha_1\delta(x-d_1) + \alpha_2\delta(x) + \alpha_3\delta(x+d_2) \right]u = 0 \tag{B1}$$

where $\alpha_j$ are the strengths of cavities. Equation (B1) possesses three localized states. We optimize the parameters of cavities $\alpha_j$ and their separations $d_j$ so that two of these states, $u_1(x)$ and $u_2(x)$ satisfy the conditions of our interest. First, we require that states $u_1(x)$ and $u_2(x)$ and, consequently, the corresponding collective states $|c_1\rangle$ and $|c_{2n}\rangle$, are coupled to a single waveguide only and maximized in the region of QEs as illustrated in Figs. 6(a) and 8(a). To this end, state $u_1(x)$ [$u_2(x)$] is designed to be finite and state $u_2(x)$ [$u_1(x)$] is designed to vanish near the input (output) waveguide [Fig. 8(c) and 8(d)]. For the dipole QE-field interaction, we have $V_{pn} \sim D_p u_p(x_n)$, where $u_p(x)$ is the normalized photon state, $x_n$ is the position of QE $n$, and $D_p$ is the dipole matrix element between states of QE [17]. In Fig. 8(c), normalized functions $u_1(x)$ and $u_2(x)$ are approximately equal in the center area (region of QEs) which corresponds to $u_1(x)/u_2(x) = 1$ and couplings $V_{1n}/V_{2n} = D_1/D_2$ independent of the position of QEs. In Fig. 8(d), the values of normalized $u_1(x)$ and $u_2(x)$ are proportional in the region of QEs with a factor of 1.7 which corresponds to $V_{1n}/V_{2n} = 1.7D_1/D_2$.

**APPENDIX C**

From Eq. (12) in the absence of losses the condition of unity inelastic transmission probability, $P = 1$ is written as

$$\left| \Delta\omega_1\left( \Delta\omega_2\Delta\omega_a - V_2^2 \right) - N\Delta\omega_2 V_1^2 \right|^2 = 4\pi^2 NW_{11}^2 W_{22}^2 V_1^2 V_2^2 \tag{C1}$$

We introduce:

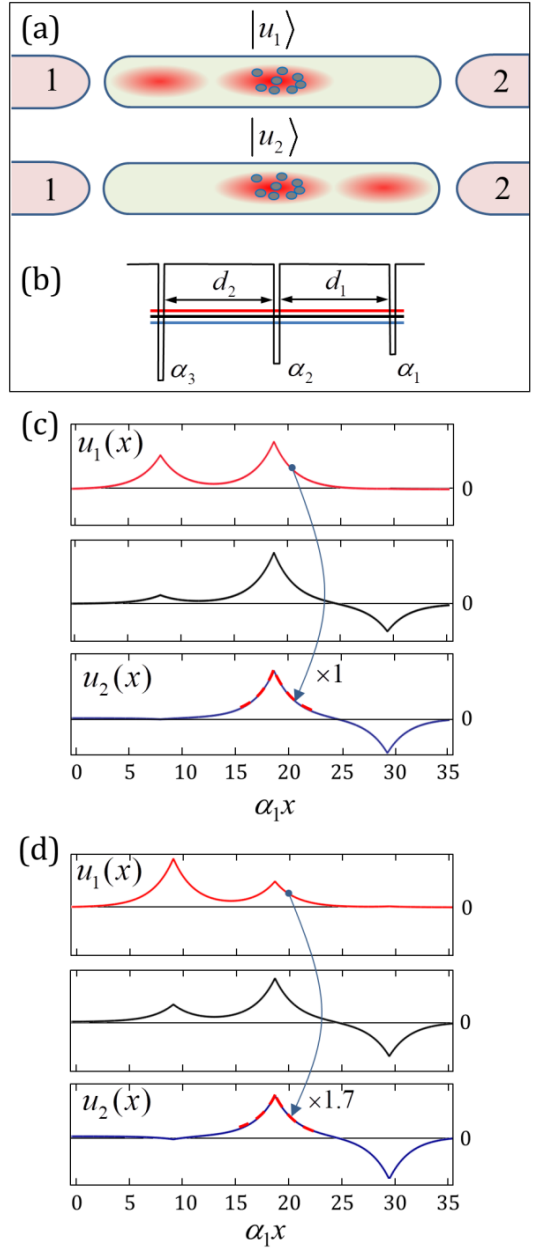

**Fig. 8.** One-dimensional optical microcavity enabling full conversion of the single photon frequency. (a) Illustration of the distribution of localized optical states and QEs. (b) Three coupled short-range cavities with strengths $\alpha_j$ and separations $d_j$ which are designed to resemble the distributions of localized states illustrated in the bottom of Fig. 6(a) and proportional to each other in the region of QEs. (c) Localized states $u_1(x)$ and $u_2(x)$ which are equal to each other in the region of QEs. (d) Localized states $u_1(x)$ and $u_2(x)$ proportional to each other in the region of QEs with the factor 1.7. The middle plot in (c, d) shows the third localized state of the microcavity which is strongly coupled to both waveguides and therefore is not appropriate for our application.

$$\Delta_1 = \omega - \omega_1^{(c)}, \qquad \gamma_1 = \pi W_{11}^2,$$
$$\Delta_2 = \omega - \omega_2^{(c)} + \omega_{g_1}^{(a)} - \omega_{g_2}^{(a)}, \qquad \gamma_2 = \pi W_{22}^2, \tag{C2}$$
$$\Delta_a = \omega - \omega_e^{(a)} + \omega_{g_1}^{(a)}.$$

Then Eq. (C1) is rewritten as:

$$\left(\Delta_1 V_2^2 + N\Delta_2 V_1^2 - \Delta_0\Delta_1\Delta_2 + \Delta_0\gamma_1\gamma_2\right)^2 + \left(\gamma_1 V_2^2 + N\gamma_2 V_1^2 - \Delta_0\Delta_1\Delta_2 - \Delta_0\Delta_2\gamma_2\right)^2 = 4N\gamma_1\gamma_2 V_1^2 V_2^2 \tag{C3}$$

To simplify this condition, we recall that the unity transmission corresponds to its maximum, i.e., to the minimum of the denominator in Eq. (12) equal to the left hand side of Eq. (C3). This minimum is determined by zeroing of the partial derivatives of the left hand side in Eq. (C3) with respect to $\Delta_0$, $\Delta_1$, and $\Delta_2$:

$$\left(V_2^2 - \Delta_0\Delta_2\right)\Omega - \Delta_0\gamma_2\Theta = 0,$$
$$\left(NV_1^2 - \Delta_0\Delta_1\right)\Omega - \Delta_0\gamma_1\Theta = 0, \tag{C4}$$
$$\left(\gamma_1\gamma_2 - \Delta_1\Delta_2\right)\Omega - \left(\Delta_1\gamma_2 - \Delta_2\gamma_1\right)\Theta = 0$$

where

$$\Omega = \Delta_1 V_2^2 + N\Delta_2 V_1^2 - \Delta_0\Delta_1\Delta_2 + \Delta_0\gamma_1\gamma_2,$$
$$\Theta = \gamma_1 V_2^2 + N\gamma_2 V_1^2 - \Delta_0\Delta_1\Delta_2 - \Delta_0\Delta_2\gamma_2. \tag{C5}$$

From these equations, we have:

$$N\gamma_1 V_1^2\left(\Delta_2^2 + \gamma_2^2\right) = \gamma_2 V_2^2\left(\Delta_1^2 + \gamma_1^2\right) \tag{C6}$$

$$\Delta_0\gamma_1\left(\Delta_2^2 + \gamma_2^2\right) = V_2^2\left(\Delta_1\gamma_2 + \Delta_2\gamma_1\right) \tag{C7}$$

Substitution of $\Delta_0$ found from Eq. (C7) into Eq. (C3) leads to Eq. (C6). Thus, satisfaction of Eqs. (C6) and (C7) presents the criterion for the inelastic transmission probability equal to unity. These equations are equivalent to Eq. (14) of the main text.